\documentclass[aps,twocolumn,noshowpacs,noshowkeys,superscriptaddress
]{revtex4-1}
\usepackage{amsfonts}
\usepackage{amsmath}
\usepackage{amssymb}
\usepackage{graphicx}
\usepackage{hyperref}

\usepackage{textcomp}%
\newcommand {\ket}[1]{|{#1}\rangle}
\newcommand{\vv}[1]{\mathbf{#1}}

\begin{document}
\def\neel{Institut N\'{e}el, Universit\'{e} Grenoble Alpes - CNRS:UPR2940, 38042 Grenoble, France}
\def\ilm{Institut Lumi\`{e}re Mati\`{e}re, UMR5306, CNRS - Universit\'{e} Claude Bernard Lyon 1, 69622 Villeurbanne, France}
\author{L. Mercier de L\'{e}pinay}
\affiliation{\neel}
\author{B. Pigeau}
\affiliation{\neel}
\author{S. Rohr}
\affiliation{\neel}
\author{A. Gloppe}
\affiliation{\neel}
\author{A. G. Kuhn}
\affiliation{\neel}
\author{P. Verlot}
\affiliation{\neel}
\author{E. Dupont-Ferrier}
\affiliation{\neel}
\author{B. Besga}
\affiliation{\neel}
\author{O. Arcizet}
\affiliation{\neel}
\email{olivier.arcizet@neel.cnrs.fr}

\title{Nano-optomechanical measurement in the photon counting regime}

\begin{abstract}
{\bf Optically measuring in the photon counting regime is a recurrent challenge in modern physics and a guarantee to develop weakly invasive probes. Here we investigate this idea on a hybrid nano-optomechanical system composed of a nanowire hybridized to a single Nitrogen-Vacancy (NV) defect. The vibrations of the nanoresonator grant a spatial degree of freedom to the quantum emitter and the photon emission event can now vary in space and time. We investigate how the nanomotion is encoded on the detected photon statistics and explore their spatio-temporal correlation properties. This allows a quantitative measurement of the vibrations of the nanomechanical oscillator at unprecedentedly low light intensities in the photon counting regime when less than one photon is detected per oscillation period, where standard detectors are dark-noise-limited. These results have implications for probing weakly interacting nanoresonators, for low temperature experiments and for investigating single moving markers.}
\end{abstract}

\maketitle

Recent developments in hybrid mechanical  quantum systems share the common objective of creating and exploring non-classical states of motion of macroscopic objects \cite{Schwab2005,Treutlein2013}. These systems combine two elementary bricks of quantum mechanics: a mechanical oscillator and a two level system, in the form of superconducting qubits \cite{LaHaye2009,O'Conell2010,Pirkkalainen2013}, single spins \cite{Arcizet2011,Kolkowitz2012,Bennett2012, Yacoby2012,Ganzhorn2013,Rohr2014,Teissier2014,Ovartchaiyapong2014,Pigeau2015}, quantum dots \cite{Lassagne2009, Steele2009,Sallen2009,Bennett2010,Yeo2013}, BEC \cite{Camerer2011,Joeckel2014}, molecules \cite{Tian2014} or ions.
Advances in this field are oriented towards both increasing the hybrid coupling strength and reducing the nanoresonator dimensions in order to maximize the qubit sensitivity to the nanoresonator dynamics. Since this reduction in size concomitantly weakens its interaction with standard opto- or electro-mechanical probe fields, it is important to investigate alternative nanomotion readout strategies, directly based on qubit state measurement. Most of the  interfaced qubits can be probed through optical or microwave fields. We analyze here how the emitted photon statistics convey information on the oscillator nanomotion.\\
\begin{figure}[b!]
\begin{center}
\includegraphics[width=0.98\linewidth]{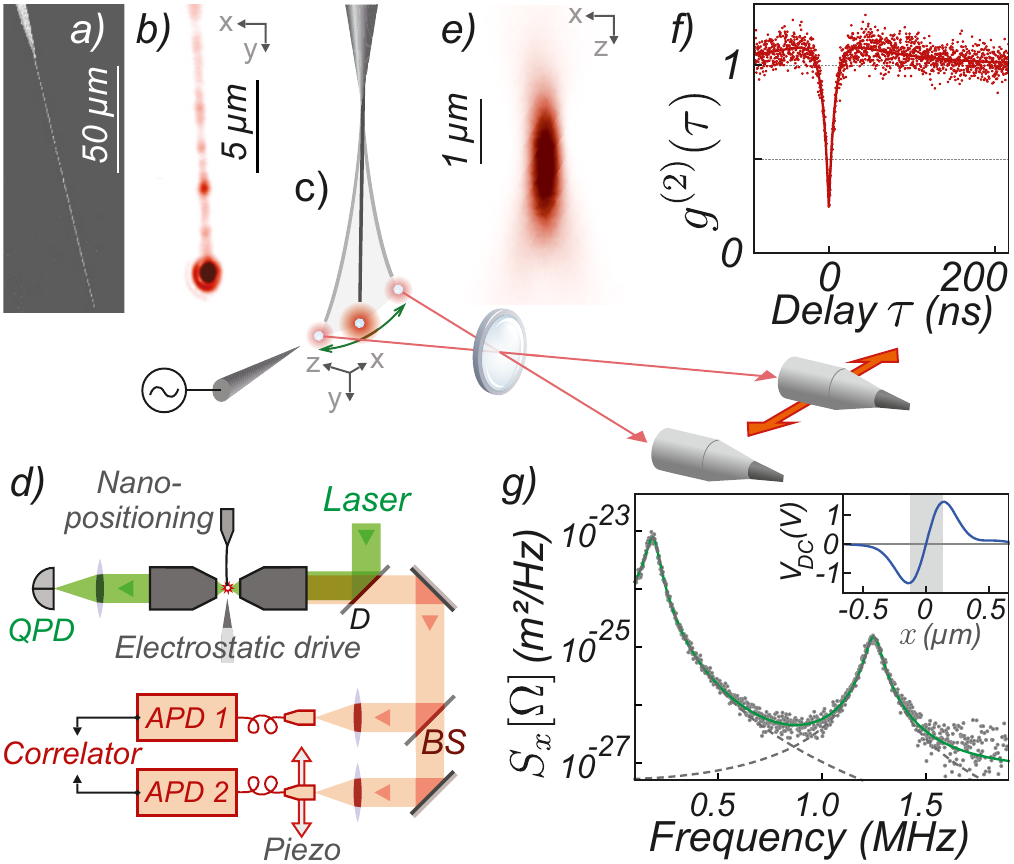}
\caption{
\textbf{ A single quantum emitter hybridized to a nanomechanical oscillator.} SEM (a) and CCD fluorescence  image (b) of the hybrid system revealing the presence of a bright quantum emitter at the nanowire extremity. (c, d): Sketch of the experimental setup (D: dichroic, BS: beam splitter, APD: avalanche photodiode, QPD: quadrant photodiode). (e): normalized fluorescence point spread function (PSF) $\Pi(\vv{r})$ of the apparatus measured by scanning the NV defect in the horizontal  xz plane. A dual single photon counter arrangement allows to verify the emitter single photon source character through intensity autocorrelation measurements (f). (g): Calibrated displacement noise spectrum of the nanowire revealing the Brownian motion of the first eigenmodes at 300\,K and atmospheric pressure \cite{Gloppe2014} (inset: differential DC transmission scan along $x$).}
\label{Fig1}%
\end{center}
\end{figure}
In this article we investigate these ideas on the example of a NV center acting as a single photon source attached to the vibrating extremity of a nanomechanical oscillator \cite{Arcizet2011,Pigeau2015}. Its nanomotion becomes encoded onto the quantum emitter state once immersed in a strongly confined pump light field, as obtained in the focus of a high numerical aperture objective. We investigate how the measured photon statistics are impacted when the position of the single emitter is moving in space due to the nanomotion. The Brownian motion of the nanoresonator is responsible for a novel photon bunching signature, consequence of the finite size of the measurement volume. Measuring the temporal cross-correlation function of the photon fluxes collected from different locations in space permits to investigate the emitter trajectory in space and time and to establish connections with the autocorrelation function of the nanoresonator position fluctuations. Finally, we illustrate the possibility of probing the thermal noise of the nanoresonator at ultralow light intensities ($\simeq$100 aW), in the photon counting regime, when photon counting rates become comparable to the oscillation frequency.\\
Various mechanisms responsible for position dependent absorption or fluorescence rates can modify the qubit emission statistics, as investigated in single trapped ion experiments \cite{Diedrich1987,Rotter2008,Bushev2013} or proposed for hybrid nanomechanical systems \cite{Puller2013,Muschik2014}. However this work focuses on fundamental variations of the detected photon statistics related to the measurement apparatus. It can thus be viewed as a dynamical extension of the original experiments of Hanbury-Brown and Twiss \cite{HanburyBrown1952} who derived the spatial coherence of a light source through static correlation measurements from different locations in space.  This work is of practical interest for nano-optomechanics at ultra-low photon fluxes, for nanosystems weakly coupled to light fields and those which cannot sustain large optical powers, such as carbon nanotubes or cryogenic environments.
Furthermore, we suggest that this approach can directly be transposed in experiments on optically trapped single quantum emitters \cite{Horowitz2012,Geiselmann2013} or for investigating the diffusion properties of biological markers.


{\it A single photon source with a mechanical degree of freedom--}
The quantum emitter, a single NV defect hosted in a $\simeq$\,50\,nm  diamond nanocrystal is attached to the free extremity of a  $46\,\rm\mu m$ long SiC nanowire with a diameter of 200\,nm,  mounted at the apex of a metallic tip (see Fig.\,1)\cite{Arcizet2011}.  The hybrid system is investigated with a confocal microscope apparatus based on high numerical aperture objectives (0.75 NA) (see SI). The 532\,nm pump laser serves for both measuring the position of the hybridized nanoresonator using the transmitted or reflected beams \cite{Gloppe2014} and pumping the NV defect. Its fluorescence in the 630 -750\,nm band is detected on avalanche photodetectors operated in the photon counting regime, featuring dark noise lower than 50 counts per second ($\simeq 10\,\rm aW$). Fluorescence images of the NV-functionalized nanoresonator reveal the presence of a NV defect at its extremity, see Fig.\,1b,  whose single photon source character is verified through autocorrelation measurements (Fig.\,1f). Piezo-scanning the suspended NV defect in the waist area allows to determine the fluorescence point-spread-function $\Pi(\vv{r})$ of the apparatus (Fig.\,1e), featuring a minimum waist of $w_0\simeq 380\,\rm nm$. The pump laser spot size can also be broadened by defocussing to produce an homogeneous illumination over the  NV oscillating trajectory, which can be efficiently driven through electrostatic actuation, see Fig.\,2.  In order to collect the fluorescence from different positions across the emitter trajectory, the detectors are mounted on piezo-positioners  (see SI). This permits to explore the photon cross-correlations from different points in space \cite{HanburyBrown1952}.\\
\begin{figure}[b]
\begin{center}
\includegraphics[width=0.98\linewidth]{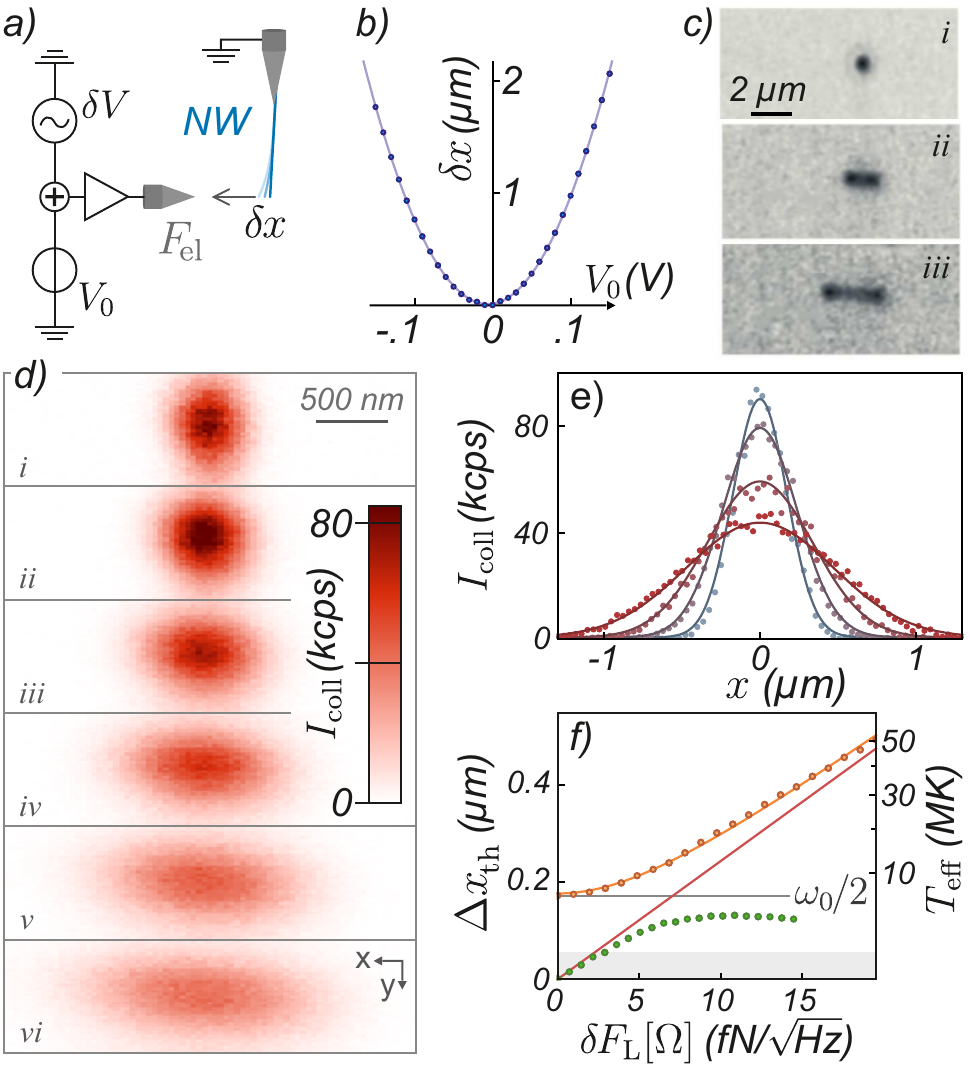}
\caption{
\textbf{Optically resolved enhanced Brownian motion.} (a) Electric scheme employed for electrostatic actuation, combining an offset $V_0$ and the drive signal $\delta V$. (b) Static deflection obtained for a $\simeq$\,10\,$\mu$m electrode-nanowire spacing, presenting a quadratic dependence in the voltage applied. The force $\delta F$ becomes linear in the (weak) control signal when added to a large offset. (c): Employing a monochromatic signal resonant with the first mechanical resonance drives a coherent oscillation of the nanowire, with a characteristic camelback shape visible in CCD fluorescence images .  (d): Using a white Gaussian voltage noise thus permits generating an additional Langevin force which increases the effective vibrational temperature $T_{\rm eff}$ of the fundamental mode.  Scanning fluorescence images in the vertical  xy plane for increasing excess noise strength which permits a direct visualization of the position distribution of the enhanced nanowire thermal noise (rms amplitude $\Delta x_{\rm th}$). Their horizontal cross sections shown in (e) are adjusted with Gaussian distributions of width $\sqrt{w_0^2/4+\Delta x_{\rm th}^2}$ (see text) and reported in (f) (solid lines). The green dots represent the results of direct noise thermometry which are only valid for spatial spreadings smaller than the optical waist (gray region).}
\label{Fig2}%
\end{center}
\end{figure}
The Brownian motion of the nanowire is detected on a quadrant photodetector  (see method in \cite{Gloppe2014}), its displacement noise spectrum $S_x[\Omega]=\int{e^{i\Omega \tau}\langle\delta x(t)\delta x(t+\tau)\rangle} d\tau$ is shown in Fig.\,1g. The first eigenmode sits 30 dB on top of a shot noise limited background for 1\,mW injected power. It oscillates at a frequency $\Omega_{\rm m}/2\pi = 190\,\rm kHz$ with a quality factor of $Q\simeq2$,  drastically limited here by air acoustic damping and a fitted adjusted effective mass of $M_{\rm eff}\simeq2\times 10^{-15}\,\rm kg$.
The nanowire can be efficiently driven into motion through an electrostatic actuation by approaching a sharp electrode polarized with a time-dependent voltage $\delta V$ in the vicinity of the nanowire extremity, perpendicularly to the optical axis (see SI). Large deflections $\delta x$ can be achieved (see Fig. 2b), with a typical efficiency of $\delta F/\delta V = 50\,\rm pN/V$ for $V_0=100\,\rm V$, expressed as an equivalent local force applied perpendicularly on the nanowire extremity. For comparison, the nanoresonator force sensitivity $\delta F_{\rm th} [\Omega]=\sqrt{2 M_{\rm eff}\Gamma_{\rm m}k_b T}$ amounts to $1.4\,\rm fN/\sqrt{Hz}$ in air.\\
{\it Optically resolved enhanced Brownian motion--}
A natural metric for resolving the trajectory of the oscillating single photon source is the absolute spatial resolution of the optical apparatus, defined via its point spread function $\Pi(\vv{r})$ (Fig.\,1g). The NV fluorescence properties become strongly position sensitive when the spatial spreading of the single photon source trajectory is comparable to the optical waist. This regime has not yet been explored in existing optomechanical experiments but will soon become relevant with the ongoing trends towards reducing the oscillator mass in particular through the use of carbon based nanoresonators and improvements of the absolute optical resolution through advanced super-resolution methods \cite{Rittweger2009}.
The large efficiency of the electrostatic actuation, combined with the extreme force sensitivity of the nanowire, permits to dramatically enhance its thermal noise by applying an additional Langevin force $\delta F_L$ emulated by a noise generator delivering a spectrally white Gaussian noise (see SI). Careful piezo-positioning of the electrostatic tip permits to dominantly drive the first eigenmode in a direction perpendicular to the optical axis (x axis), the 1D trajectory being described by $\xi(t)$. Increasing the r.m.s. amplitude of the noise source thus permits enhancing the effective temperature $T_{\rm eff}$ of the first eigenmode and thus the spatial spreading  $\Delta x_{\rm th}=\sqrt{k_B T_{\rm eff}/M\Omega_{\rm m}^2}$  of the nanowire position fluctuations. This was verified at small temperature increase (Fig.\,2f) in the direct optical readout based on real-time position sensing (Fig.\,1g). At larger effective temperature, the spatial distribution of the nanowire thermal noise can exceed the transverse PSF size and can thus directly be measured through a scanning fluorescence imaging (see Fig.\,2d). The data are fitted by a convolution between the stationary Gaussian probability distribution $P(x)=\left(2\pi \Delta x_{\rm th}^2\right)^{-1/2}\exp({-x^2/2\Delta x_{\rm th}^2})$ associated with Brownian motion and the fluorescence PSF of the confocal microscope sampled across the vibration axis. The latter which is measured in absence of drive (see Fig.\,1e) is approximated at the waist by a Gaussian function $\Pi(x)= \exp({-2 x^2/w_0^2})$.  The resulting convolved Gaussian spatial width $\sqrt{{w_0^2}/{4}+\Delta x_{\rm th}^2}$ (Fig. 2f) allows measuring the spatial spreading of the nanowire thermal noise. The linearity of the actuation is verified up to extremely large rms amplitudes, approaching the $\rm \mu m$ range. This corresponds to  an equivalent effective temperatures around  $10^7\,\rm K$ for the first eigenmode and permits exploring nano-optomechanical systems far beyond the (thermal) Lamb-Dicke regime ($\theta_x\equiv\Delta x_{\rm th}/ w_0<1$).\\
{\it Autocorrelation function--}
The impact of the enhanced nanomotion on the detected photon statistics was then investigated through fluorescence second order correlation measurement from different points in space and time, $\mathrm{g}^{(2)}\left(\tau,x_{1},x_{2}\right)$. This quantity reflects the probability of detecting a photon in the stop channel,  collecting the fluorescence around a position $x_2$, at a time difference $\tau$ after having detected a photon in the start channel, centered at $x_1$.
\begin{figure}[t]
\begin{center}
\includegraphics[width=0.98\linewidth]{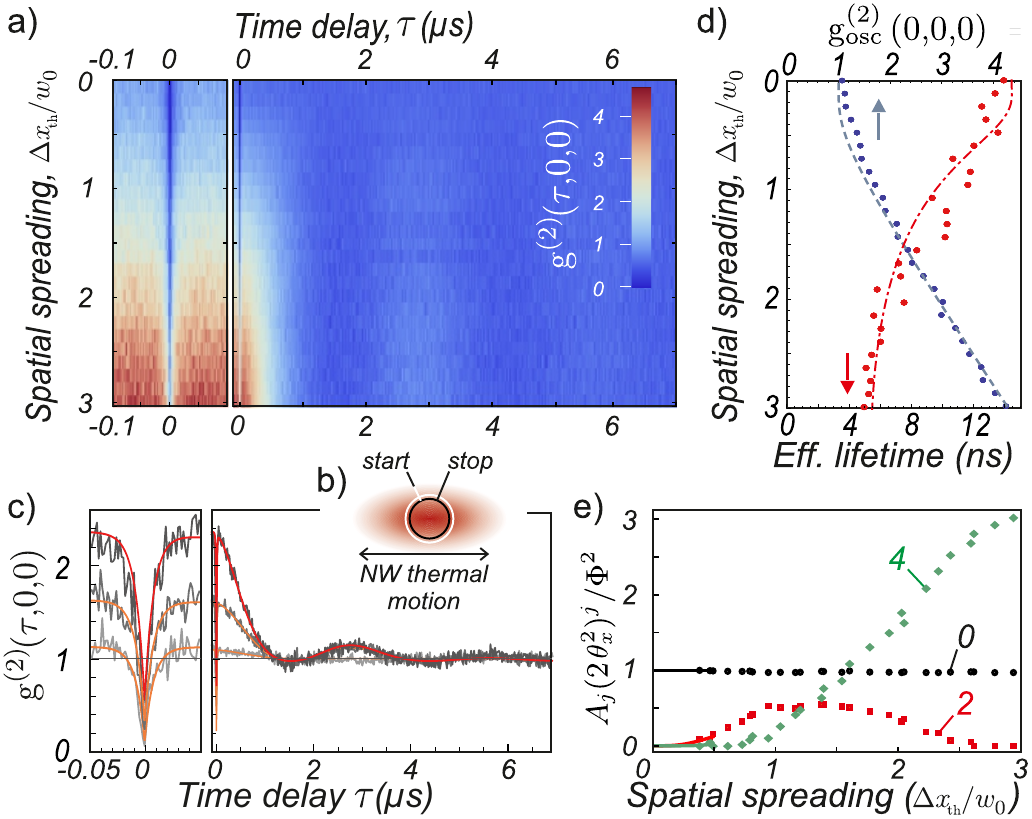}
\caption{
\textbf{Cross-correlation function of the oscillating single photon source}  (a) measured when both detectors are placed at the center of the Brownian trajectory (b) for an increasing incoherent drive amplitude and normalized using the measured photon fluxes product $\Phi^2$ \cite{Beveratos2002}.The strength of the photon bunching appearing at large excursions is reproduced in (d) and fitted with theoretical expression (see SI), as well as the reduction of the emitter effective lifetime. (c): data are well fitted with expression (\ref{eq-aj}), the fitting coefficients $A_{2j} (2\mu\theta_x^2)^{2j}/\Phi^2$ are shown in (e) and compared to the theoretical expansion (lines) converging at moderate spatial spreading (see SI).}
\label{Fig3}
\end{center}
\end{figure}
The detection of a start event at a time $\tau_1$  initializes the NV in its ground state and localizes it at the start position $\xi (\tau_1)$ near $x_1$. Its subsequent fluorescence rate is proportional to the evolution of the excited state population $\sigma_e (\tau, \xi(\tau_1+\tau))$, the photons being emitted from the position  $\xi(\tau_1+\tau)$. It can be numerically computed once the trajectory $\xi(t)$  and the pump intensity profile $I(x)$ are known.  In our numerical simulations, a simplified 3-level scheme was adopted to model the NV fluorescence properties, see SI, with a position dependent pumping rate proportional to the time-varying intensity seen by the nanowire $I(\xi(t))$. The PSF of both measurement channels were measured experimentally and can be modeled by displaced Gaussian profiles $\Pi_{i}(x)\equiv\Pi(x-x_i)$ so that the detection rates of the start,stop photons are weighted by $\Pi_{1}(\xi(\tau_1))$, $\Pi_{2}(\xi(\tau_1+\tau))$. Here we restrict ourselves to the situation where the emitter lifetime ($\Gamma^{-1}$) remains short compared to the duration of illumination, $\Delta x_{\rm th}/w_0\ll \Gamma/\Omega_{\rm m}$, so that the probability to detect a start photon  can be approximated to the local time averaged fluorescence rate at the start position, proportional to $\bar{\sigma}_e(\xi(\tau_1))$, which is directly measured in Fig.\,2d.
Finally the measured normalized spatio-temporal cross-correlation function of the vibrating single photon source, can be expressed as:
\begin{equation}\mathrm{g}^{(2)}\left(\tau,x_{1},x_{2}\right)=
\mathrm{G}^{(2)}\left(\tau,x_{1},x_{2}\right)/\mathcal{N},\label{eq-g2}
\end{equation}
where,  using $\langle\ldots\rangle_T$  as the average over the integration time $T$,  we have $\mathrm{G}^{(2)}\left(\tau,x_{1},x_{2}\right)\equiv \left\langle \bar{\sigma}_e(\xi(\tau_1)) \,\sigma_e(\tau, \xi(\tau_1+\tau)) \Pi_{1}(\xi(\tau_1)) \Pi_{2}(\xi(\tau_1+\tau)) \right\rangle_T $  which is integrated over all the possible start events ($\tau_1$) for a given trajectory $\xi(t)$.
The denominator reads $\mathcal{N}\equiv\langle \bar{\sigma}_e(\xi(t))\Pi_1(\xi(t)) \rangle_T \langle\bar{\sigma_e}(\xi(t)) \Pi_2(\xi(t))\rangle_T$ and using the ergodic principle, we have $\mathcal{N}=\Phi_1\Phi_2$ with $\Phi_i=\int{dx P(x) \Pi_i(x)}$ which normalizes to the product of the time averaged photon flux seen by each detector \cite{Beveratos2002}. These expressions have a general reach, they are valid for any trajectory in space and can account for optical saturation or for sharp optical illumination ($w_0 < \Delta x_{\rm th}$). It is interesting to note that in the particular case of broad illumination, the excited state population recovery becomes position insensitive, $\sigma_e(\tau, \xi(\tau_1+\tau))\rightarrow\sigma_e(\tau)$ so that the autocorrelation function  can be factorized as:
$
\mathrm{G}^{(2)}(\tau, x_1, x_2)= \sigma_e(\tau)\, \mathrm{G}^{(2)}_{\rm osc} (\tau, x_1,x_2),
$ with
$\mathrm{G}^{(2)}_{\rm osc}\left(\tau,x_{1},x_{2}\right)\equiv\int{d\tau_1 \Pi_1(\xi(\tau_1))}\Pi_2(\xi(\tau_1+\tau))$ capturing all of the spatial dependency. This permits to clearly isolate the measurement-related contributions to the correlation function. In the following, we will explore different experimental configurations that permit a full characterization of the emitter trajectory in space.\\
First, Fig. 3a represents $\mathrm{g}^{(2)}\left(\tau,0,0\right)$ measured when both detectors are collecting the fluorescence from the center of the single photon source Brownian trajectory.  When increasing its spatial spreading beyond the optical resolution, a pronounced oscillatory bunching signature appears, whose amplitude increases with $T_{\rm eff}$. Its magnitude is reported in Fig.\,3d and can be well fitted with $\mathrm{g}^{(2)}_{\rm osc}\left(0,0,0\right)=\int{dx P(x)\Pi(x)^2}/\Phi^2=(1+4\theta_x^2)/\sqrt{1+8\theta_x^2}$ (see SI). This reveals the emergence of a new photon loss channel for the detection path, due to photon emission outside of the detection volume. In analogy with photon bunching signatures appearing for quantum emitters presenting a dark metastable state, this bunching signature is accompanied by a sharpening of the anti-bunching profile so that the slope at origin is steepened from $14\,\rm ns^{-1}$ to $4\,\rm ns^{-1}$ at large effective temperatures. These signatures represent a quantitative analysis tool for super-resolution experiments on spatially moving targets. \\
The demonstration of this randomization of the photon emission in space due to the nanomotion was  further investigated by measuring the fluorescence cross-correlations from different locations in space, using the movable photon counting areas. Figures 4a,b represent maps of $\mathrm{G}^{(2)}\left(\tau,0,\delta\right)/\mathrm{G}^{(2)}\left(\infty,0,0\right)$, where the start counter remains centered, the stop counter being displaced across the NV trajectory. Experimentally, it is obtained by accumulating the correlations during a fixed duration $T$ at different position of the stop detector. The anti-bunching character of the single photon source is preserved all across the space direction while the bunching magnitude now varies in space and time, revealing the diffusion of the single photon source following its localization in $x_1=0$ at $\tau=0$ by the start photon detection event. Due to the reduced quality factor limited by air damping, only few oscillations are visible before erasure of the initial conditions, when the autocorrelation function loses its time dependency and converges towards a Gaussian thermal distribution (see SI). These measurements permitting to visualize the diffusion in space and time of the single emitter are performed in the photon counting regime and reproduce the results derived from a record of $\xi(t)$  in real time on the QPD obtained at lower effective temperature (see SI). Our results are also in good agreement, see Fig.\,4a,b with numerical simulations based on equation (\ref{eq-g2}), where the Brownian motion trajectory was numerically generated using the experimentally determined parameters (effective temperature, frequency, quality factor) (see SI).

\begin{figure}[b]
\begin{center}
\includegraphics[width=0.98\linewidth]{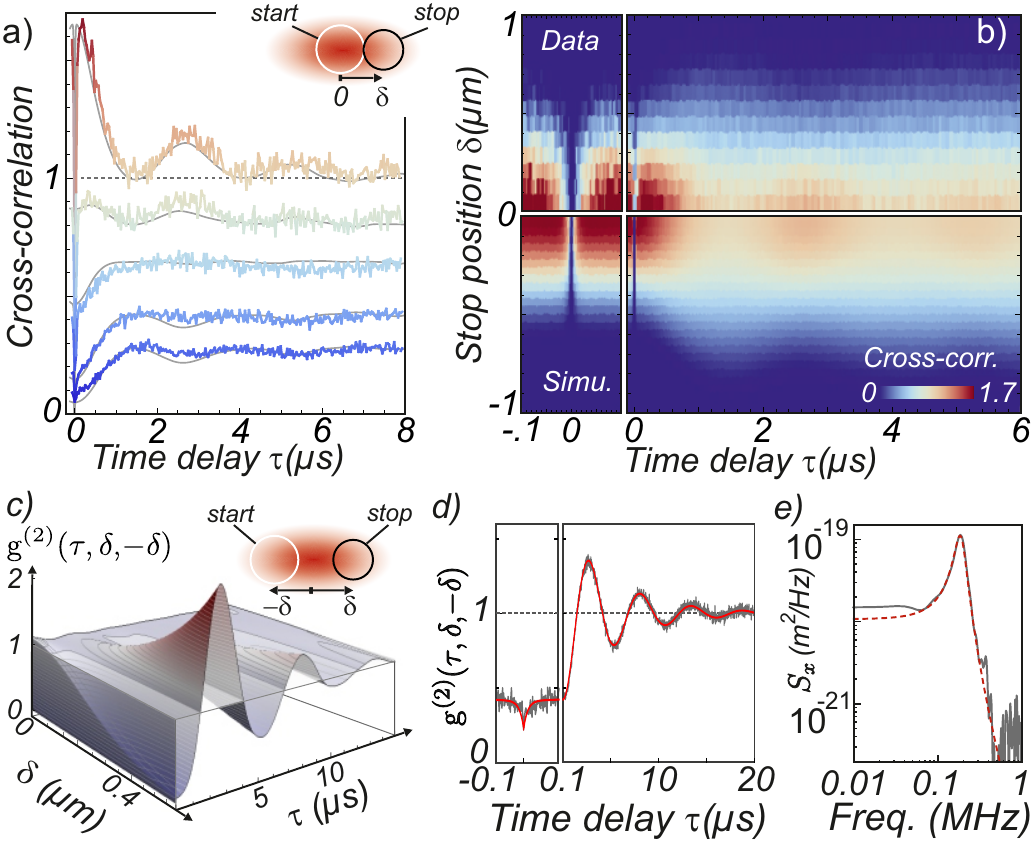}
\caption{\textbf{Probing the oscillator dynamics with a single photon source.}
Diffusion in space and time of the oscillating single photon source:  (a, b): the cross-correlation functions $\mathrm{G}^{(2)}\left(\tau,0,\delta\right)/\mathrm{G}^{(2)}\left(\infty,0,0\right)$ are measured for different positions of the stop detection area ($\delta=0,\,0.2,\,0.35,\,0.5,\,0.6\,\rm \mu m$), scanned across the spatial distribution of an enhanced Brownian motion of $0.95\, w_0$ rms spreading.  Using this normalization permits revealing signal spreading in space.  The results of numerical simulations are shown for comparison (a: full lines, b: bottom).
Measurement of the oscillator autocorrelation function:
c) results of the numerical simulation (see SI) showing the cross-correlation function $\mathrm{g}^{(2)}\left(\tau,\delta,-\delta\right)$ obtained for increasing separations $\pm\delta$ between the start/stop detection areas  (see inset) and for $\Delta x_{\rm th}/\omega_0=0.4$. d):  corresponding experimental data obtained for similar excitation strength, revealing the oscillator autocorrelation function. The data are fitted with eq. (\ref{eq-aj}). (e): Fourier transform of $g^2(\tau)-1$ fitted with a thermal noise spectrum (dashed line).}
\label{Fig1}%
\end{center}
\end{figure}

{\it Nanomotion sensing in the photon counting regime--}
Such cross-correlation measurements in space and time can be employed to measure the vibration noise spectrum of the nanowire defined as the Fourier Transform of its position autocorrelation function $C_\xi(\tau)=\langle\xi(t)\xi(t+\tau)\rangle$. Here we illustrate this connection between both correlation functions allowing to measure the thermal noise of the nanomechanical oscillators at ultralow optical intensities, falling far below the dark noise of standard photodiodes.
Exploiting the Gaussian character of the nanowire thermal noise position distribution which has been verified experimentally (Fig.\,2e), it is possible (see SI) in the case of spatially broadened illumination to expand the autocorrelation function in powers of $C_\xi(\tau)$:
\begin{equation}g^{(2)}\left(\tau,x_{1},x_{2}\right)=\frac{\sigma_e(\tau)}{A_0}\sum_{j=0}^{\infty} A_j \left(\frac{C_\xi(\tau)}{w_0^2/2}\right)^j.\label{eq-aj}
\end{equation}
The $A_j$ coefficients depend on the spatial gradients of the displaced PSF and on the thermal noise spreading $\Delta x_{\rm th}$ (see SI). $A_0$ reflects the product of the mean photon fluxes seen by each detector and thus conveys the role of the shot noise in that measurement. Maximizing $A_1$ with respect to higher orders terms allows to optimize the measurement of the nanowire autocorrelation function $C_\xi(\tau)$, which is not possible when one detector remains centered (such as in Fig.\, 3,\,4a,b)  since $A_{2j+1}= 0$ there.
Instead, it is interesting to reproduce a quadrant photodiode configuration, when both photon counters are arranged to monitor the fluorescence on each side ($\pm\delta $) of the nanowire trajectory (see Fig. 4c). In that situation, larger contrasts can be obtained for sufficient separation (see Fig.\,4c), and the quadratic term $A_2$ can even be nulled for $\delta^\star \equiv w_0/2$ (see SI). The NV cross-correlation function now presents an oscillating and exponentially decaying pattern, see Fig.\,4d, which can be perfectly adjusted using equation (\ref{eq-aj}) and the position autocorrelation function $C_\xi(\tau)=\Delta x_{\rm th}^2\, e^{-\Gamma_{\rm m}\tau/2}\left(\cos \tilde{\Omega}_{\rm m}\tau + \Gamma_{\rm m}/2\tilde{\Omega}_{\rm m}\sin \tilde{\Omega}_{\rm m}\tau\right)$, with ${\tilde{\Omega}_{\rm m}}^2\equiv\Omega_{\rm m}^2-\Gamma_{\rm m}^2/2$ \cite{Chang1945}. Its Fourier transform (Fig.\,4e) provides a measurement of the enhanced thermal noise spectrum  $S_x[\Omega]$ of the nanowire. The emitter finite lifetime is responsible for an upper bound  of $\simeq40\rm \,MHz$ on the measurement bandwidth which does not limit our measurements here.
For small excitation amplitudes ($\Delta x_{\rm th}\ll w_0$), one obtains: $g^{(2)}(\tau,\delta^\star,-\delta^\star)\approx\sigma_e(\tau)\left(1-4C_\xi(\tau)/w_0^2\right)$, (see SI) with a shot noise limited sensitivity of $w_0^2/\left(4\Phi \sqrt{\tau_{\rm bin}}\right)$ amounting to $ (6\,\rm nm)^2/\sqrt{Hz}$ for a $10^6\,\rm Hz$ mean photon flux and a bin time of $\tau_{\rm bin}=1\,\rm\mu s$. This sensitivity is sufficient in principle to detect the thermal noise of ultralight oscillators such as suspended carbon nanotubes, even at low temperatures (see SI) \cite{MoserJ.2013}.\\
We note here that it is generally possible to realize a linear measurement of the oscillator position $\xi(t)$ by recording and subtracting the photon fluxes seen by each detector arranged in a quadrant photodiode configuration. In that case, the thermal noise can be detected above the photon shot noise if $\Phi\gtrsim\Gamma_{\rm m}(w_0/\Delta x_{\rm th})^2$ which requires having large photon fluxes, light oscillators or high mechanical quality factors. However this method suffers important experimental constraints since in order to convert the photon counting signals into a meaningful time-resolved intensity signal, it is necessary to choose an integration time comparable to the inverse mean photon rate  ($\Phi^{-1}$). In order to probe the nanomotion in real time, this integration time should be smaller than the oscillation period, so that it requires in turn to detect at least one photon per oscillation period $\Phi>\Omega_{\rm m}$. In contrast, our experiment operates below this criterion, which underlines the strength of this method based on second order fluorescence cross-correlations, even for detecting high frequency oscillators. \\

{\it Conclusions--}
This proof-of-principle experiment demonstrates the possibility to detect and analyze the vibrations of nanomechanical oscillators in the photon counting regime by recording the spatial and temporal correlation functions of the emitted photon flux. The detected optical powers involved fall in the sub-fW range, where standard analog detectors are dark-noise-limited. Furthermore this permits to operate at photon fluxes smaller than the oscillation frequency, where time-resolved position measurements are not possible with analog sensors. Therefore our approach also proves useful when very limited optical powers must be employed, such as in cryogenic experiments. The exposed measurements can virtually be transposed to any nanomechanical system that is weakly coupled to light (such as nanotubes) by working on Raman scattered photons \cite{Reserbat-Plantey2012} or on other defect-related fluorescence signals \cite{Hoegele2008, Wilson-Rae2012,Reserbat-Plantey2012} but also to optically trapped emitters \cite{Geiselmann2013,Kiesel2013}. The sensitivity of the measurement can be significantly improved through the use of optical super-resolution techniques \cite{Rittweger2009} or microwave assisted sharpening of the spin fluorescence PSF \cite{Grinolds2014,Maurer2010}. The investigation of the qubit dynamics second order correlation function can also be extended to the frequency domain of parametrically coupled hybrid mechanical systems, through spectral diffusion analysis \cite{Sallen2010}, or resonant optical pumping \cite{Wilson-Rae2004, Bushev2013,Puller2013,Pigeau2015}, where larger coupling strengths can be achieved.\\

\textit{Acknowledgements---} We thank V. Jacques, J.-F.\,Roch, J.-P.\,  Poizat, G.\, Bachelier, F.\, Pistolesi, P.\,Vincent, P.\,Poncharal, J.\,Jarreau, C.\,Hoarau, E.\,Eyraud and D.\,Lepoittevin, for theoretical, experimental and technical assistance. This project is supported by ANR (RPDoc-2010 and FOCUS), the ERC Starting Grant StG-2012-HQ-NOM and Lanef (CryOptics). S.R. acknowledges funding from the Nanoscience Foundation.

%

\widetext
\clearpage
\begin{center}
\textbf{\Large Supplementary material for ``Nano-optomechanical measurement in the photon counting regime"}
\end{center}

\setcounter{equation}{0}
\setcounter{figure}{0}
\setcounter{table}{0}
\setcounter{page}{1}
\makeatletter
\renewcommand{\theequation}{S\arabic{equation}}
\renewcommand{\thefigure}{S\arabic{figure}}
\renewcommand{\bibnumfmt}[1]{[S#1]}
\renewcommand{\citenumfont}[1]{S#1}

\begin{flushleft}
{\large \bf Experimental setup}
\end{flushleft}

\begin{figure}[ht!]
\begin{center}
\includegraphics[width=0.9\linewidth]{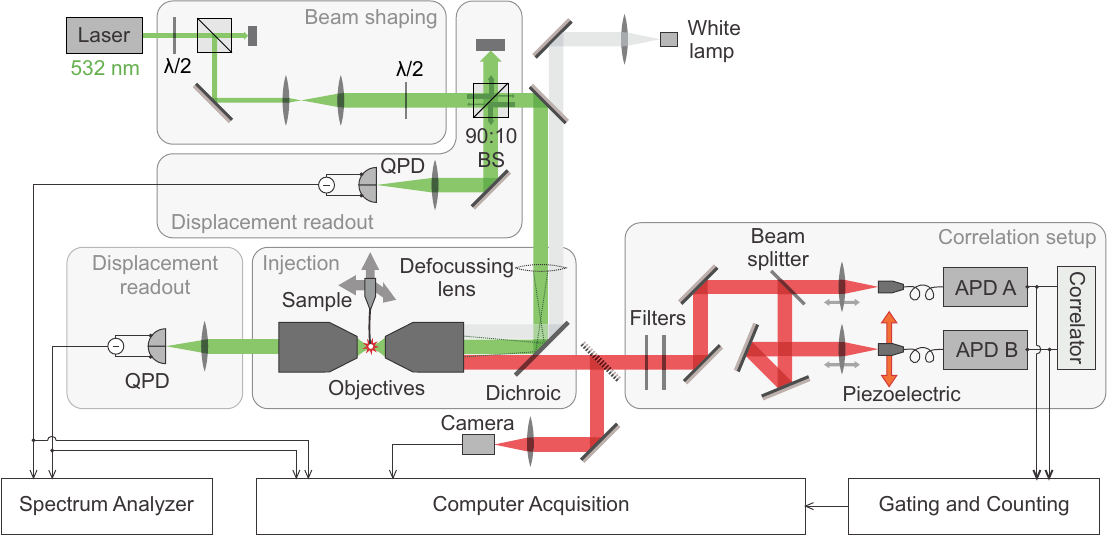}
\caption{Simplified scheme of the experimental setup. BS: Beam Splitter, QPD: Quadrant Photodiode, APD: Avalanche PhotoDiode (single photon counting modules).}
\label{fig:setupSI}
\end{center}
\end{figure}

\begin{figure}[ht!]
\begin{center}
\includegraphics[width=0.5\linewidth]{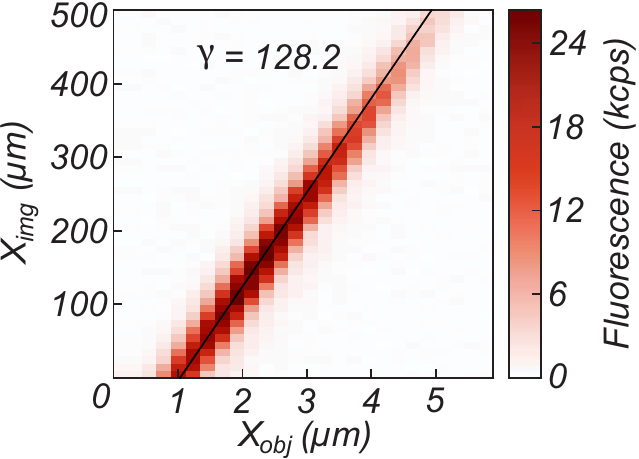}
\caption{Calibration of the real magnification factor. Fluorescence map obtained by scanning  the object and movable detector positions. The laser light field is broadened by a defocussing lens placed before the objective, which enables the NV center to be lit over several microns. The solid black fitted line yields a magnification factor of 128.2, in good agreement with the expected value 125 for the combination of focalization lens and objective used here.}
\label{fig:cal_magn2}
\end{center}
\end{figure}

This paragraph gives a more detailed description of the setup shown in \ref{fig:setupSI}.
The laser is a Laser Quantum ``Gem" laser ($532\,\rm nm$). The optical power, size and polarization of the beam can be adjusted prior to its injection in the microscope objective. The optical power is controlled with a combination of a multi-order half-wave plate and a polarizing beam-splitter. A telescope is used to match the beam diameter with the size of the input iris. Another half-wave plate is used for polarization control.\\
The beam is then focused in a long working-distance ($4\,\rm mm$) and high numerical aperture Zeiss microscope objective (x100, NA=$0.75$). After interaction with the functionalized nanowire, some of the back-scattered light is collected through the same objective.\\
A Semrock $580\,\rm nm$ dichroic mirror serves to separate the intense green excitation beam (about $0.5$ to $3\,\rm mW$) from the weak red fluorescence outcome (about $1\,\rm fW$) from the sample. This dichroic mirror stands just in front of the microscope objective and reflects the green laser into the objective while it lets the red fluorescence pass through and reach the fluorescence analysis stage. The backscattered green light is also sent back via the same optical path as the injected laser to be analysed. A third collection path is set up by an identical microscope objective positioned opposite the injection objective to collect the transmitted green light or the fluorescence.\\
A white light source is also injected through the backside of one of the injection mirrors. Its reflection on the sample is collected through the dichroic mirror and directed with a mirror to a Watec 910HX camera which permits fluorescence imaging and sample monitoring.\\
Additional filters are added after the dichroic mirror in order to completely suppress the pump light in the fluorescence analysis path. It is split with a non-polarizing beam-splitter into two ideally equally intense optical paths fed into two Avalanche PhotoDiodes (Single Photon Counting Modules) through two $50\,\rm \mu m$ core multimode fibers. The fluorescence signal is converted into voltage pulses by the APDs which are in turn either gated and counted by the interface program to perform nanowire positioning and tracking or fed to a FastComTec MCS6A correlator to build intensity correlations.
One of the fiber couplers is mounted onto a PI $500\,\rm\mu m$ range lateral piezo axis so as to move it in the image plane of the sample and to build intensity correlations from different points in space. This supplementary axis was first used to determine the real magnification factor of the experiment (\ref{fig:cal_magn2}).

An asymmetric  (90:10) beam splitter reflects most of the reflected light towards a Quadrant-PhotoDiode (QPD) with a homemade differential amplifier. The difference and the sum of the voltages of the two quadrants of the photodiode are measured and both the DC and HF components are split, thus resulting in four output channels. The HF component is fed into a Spectrum Analyzer (Agilent MXA) to build the position correlation of the nanowire and the DC signal is fed to the interface program for imaging purposes.\\

The mechanical oscillator is a $46\,\rm nm$-long silicon carbide nanowire attached to an electrochemically etched tungsten tip. The etching, preliminary characterization and binding of the nanowire to the tip are performed in the ILM in Lyon in collaboration with A. Siria, P. Vincent, A. Ayari and P. Poncharal. A XYZ Piezostage supporting the sample holder of the nanowire can be moved in front of the microscope objectives so as to scan maps of the reflected, transmitted and fluorescence lights as a function of the nanowire position in the optical beam.  Generally, two axes are scanned to produce "XY" $30\times 30\,\rm \mu m$ or "XZ" $30\times 10\,\rm \mu m$ maps. These scans and the analysis of all signals are performed thanks to a NI e-6323 measurement card and a homemade Python-Qt interface adapted to the setup. To get the laser waist to lie in the  $30\times 30\times 10\,\rm \mu m$ piezo window, the whole piezostage is mounted on a XYZ Newport translation.\\
Nanodiamonds are attached to the nanowire by dipping the nanowire into a droplet of commercial solution containing $50$ to $100\,\rm nm$ diameter nanodiamonds. They are small enough to contain 0 or 1 defect at most, given the typical NV defect densities of this type of diamond.

\begin{figure}[ht!]
\begin{center}
\includegraphics[width=0.7 \linewidth]{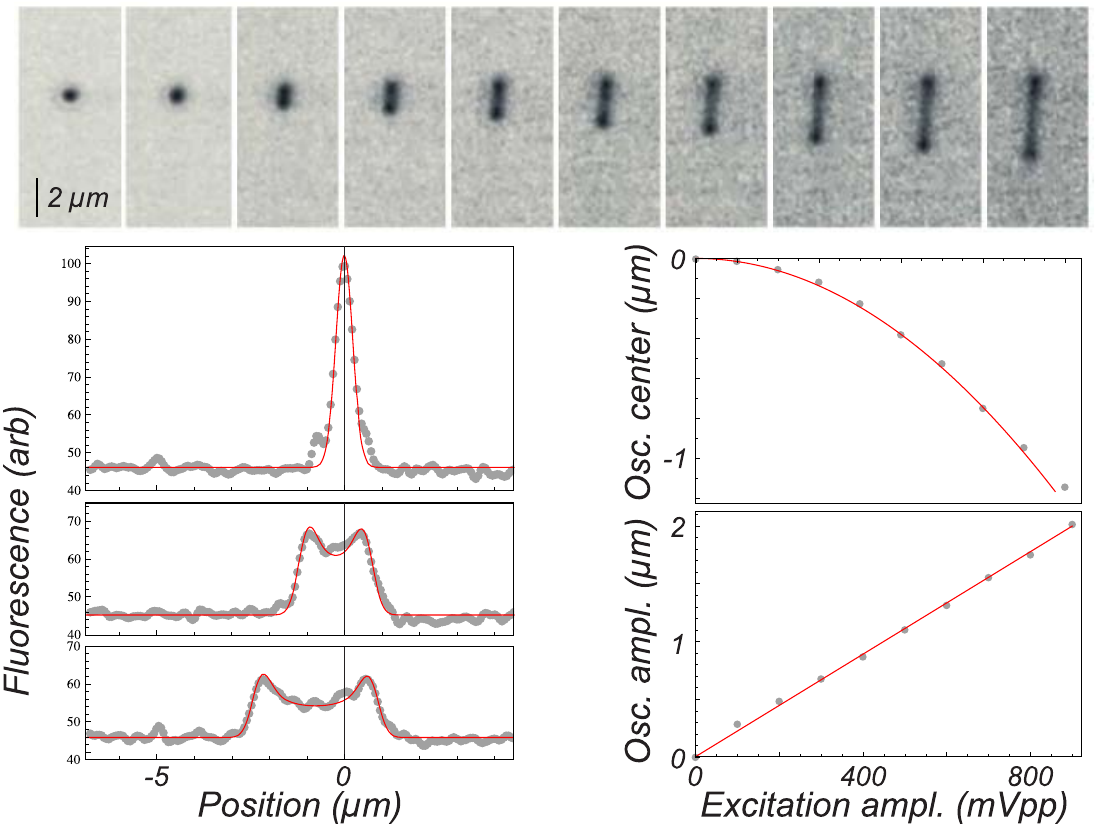}
\caption{ CCD fluorescence images obtained under broadened spatial illumination for increasing coherent driving of the nanowire. The Airy peak (left) observed at rest is progressively broadened, presenting a characteristic camel back shape. The driven trajectory can spread over several $\mu m$. To do so a resonant RF tone is applied on the electrostatic tip, with amplitude from 0 to 45 V (after the voltage amplifier)  and offset of 20V. The profiles are fitted with an oscillating Gaussian function: $F(x)=\frac{1}{T_{\rm m}}\int_{0}^{\rm T_m}{ e^{- 2 (x- x_0-\Delta x \cos \Omega_{\rm m} t )^2/w_0^2} dt }$ averaged over one mechanical oscillation (period $T_{\rm m}$) centered in $x_0$ with an amplitude $\Delta x$. These two quantities are reported for increasing drive amplitudes and fitted with quadratic an linear functions respectively.}
\label{fig:cal_magn}
\end{center}
\end{figure}

\begin{flushleft}
{\large \bf Mechanical vibrations}
\end{flushleft}

The vibrations of the nanowire are probed with a focused laser beam illuminating the NV defect, which is located at the nanowire extremity. The description of the optical readout of the deformations of the nanowire is presented in \cite{Gloppe2014}, where the multi-modal expansion is presented in detail. It permits defining an effective mass, see \cite{Pinard1999}, which depends
on the vibrational mode considered  and on the position along the nanowire. In case of a small optical waist with respect to the length $L$ of the nanowire (and to the other characteristic sizes of the mode deformation profile) the optical readout permits a local measurement of the nanowire deflection. The effective mass can be expressed as:
\begin{equation}
M_{\text{eff}, n}=M\int_{0}^{L}\frac{dy}{L}\frac{{u_{n}}^{2}(y)}{{u_{n}}^{2}(L)},
\end{equation}
where $u_{n}(y)$ is the displacement of the beam at longitudinal position $y$. $y=L$ corresponds to the free extremity and $y=0$ the clamped extremity of the nanowire. The numerical resolution of the Euler-Bernoulli equation for a singly clamped beam gives the non-normalized shape of the spatial modes of the nanowire (see Table \ref{table:spmode}), which in turn allows to calculate the effective mass of each vibrational mode (see Table \ref{table:spmode}). We have:
\begin{equation}
u_{n}(t)=\bigg\{ \Big[ \cos(k_{n} y)-\cosh(k_{n} y)\Big]+ A_{n}\Big[\sin(k_{n} y)-\sinh(k_{n} y)\Big]\bigg\}.
\label{eq:eqmodes}
\end{equation}

\begin{table}[ht]
\begin{center}
\begin{tabular}{|c|c|c|c|}
\hline
$n$ & $k_{n}L$& $A_{n}$ &$M_{\rm eff}/M$\\
\hline
1   &  1.87510  & -0.7341 & 0.2500\\
2   &  4.69409 & -1.0185 & 0.2500\\
3   &  7.85476  & -0.9992 & 0.2433 \\
4   &  10.9955  & -1.0000 & 0.9547\\
5   &  14.1372   & -1.0000  & 0.9646 \\
\hline
\end{tabular}
\end{center}
\caption{Numerical coefficients corresponding to the first 5 eigenmodes of equation \ref{eq:eqmodes}, in case of a point-like optical measurement (homogeneous mode deformation profile within the waist area). }
 \label{table:spmode}
\end{table}

This work has been realized at ambient pressure, where the two orthogonal polarizations of each longitudinal eigenmode family are not resolved due to air damping. In addition,  the nanowire is always positioned on the optical axis, where the quadrant photodiodes are only sensitive to the nano-motion perpendicular to the optical axis. Furthermore the electrode is micro-positioned perpendicularly to the optical axis, so that it generates a force that is driving the nanowire perpendicularly to the optical axis. All these reasons justify the simplification of the description of the nanowire of the nanowire vibrations by a uniaxial oscillator, only oscillating perpendicularly to the optical axis. A full description of the nanowire dynamics in 3D can however be found in \cite{Gloppe2014}, but is not required here to fully describe our results. In the following and in the article, we then describe the measured transverse position of the nanowire by the scalar $\delta x(t)$.\\
Furthermore, we will only restrict  our study to the fundamental eigenmode. This is fully justified since the second longitudinal eigenmode oscillates around 6 times faster than the fundamental period. This means that compared to the fundamental mode, the r.m.s. Brownian motion spreading of the second mode at the extremity is around 36 times smaller and the noise power of thermal noise ($\propto Q/M/\Omega_{\rm m}^3$) 216 times smaller at resonance. Moreover, the spatial profile of the electrostatic actuation is dominantly exerted at the extremity of the nanowire, in front of the electrostatic tip, so that the relevant increase of the effective temperature of the higher order modes is significantly reduced as compared to its effect on the fundamental eigenmode. This is even more pronounced due to the finite bandwidth of the noise generator that can be tuned to fade at higher frequencies, without impacting on the white character of the noise seen by the fundamental mode.
The mechanical susceptibility $\chi[\Omega]$ is defined as:
\begin{equation}
\delta x[\Omega]=\chi[\Omega]\delta F[\Omega],
\end{equation}
where $\delta x[\Omega]$ is the spectral component of the displacement measured at the free extremity of the nanowire and $\delta F[\Omega]$ is the spectral component of an external force.  Following the previously mentioned approximations, the nanowire dynamics can be  assimilated to a single mode oscillator with the susceptibility:
\begin{equation}
\chi[\Omega]=\frac{1/M_{\rm eff}}{(\Omega_{\rm m})^{2}-\Omega^{2}-i\Gamma_{\rm m}\Omega_{\rm m}}.
\end{equation}
The mechanical displacement spectrum is defined as:
\begin{equation}
S_{x}[\Omega]=|\chi[\Omega]|^{2}S_{\rm F}[\Omega],
\end{equation}
with
\begin{equation}
S_{\rm F}[\Omega]= 2 M_{\rm eff} \Gamma_{\rm m} k_B T,
\end{equation}
where T is the effective vibrational temperature of the nanowire.\\

\begin{figure}[ht!]
\begin{center}
\includegraphics[width=10cm]{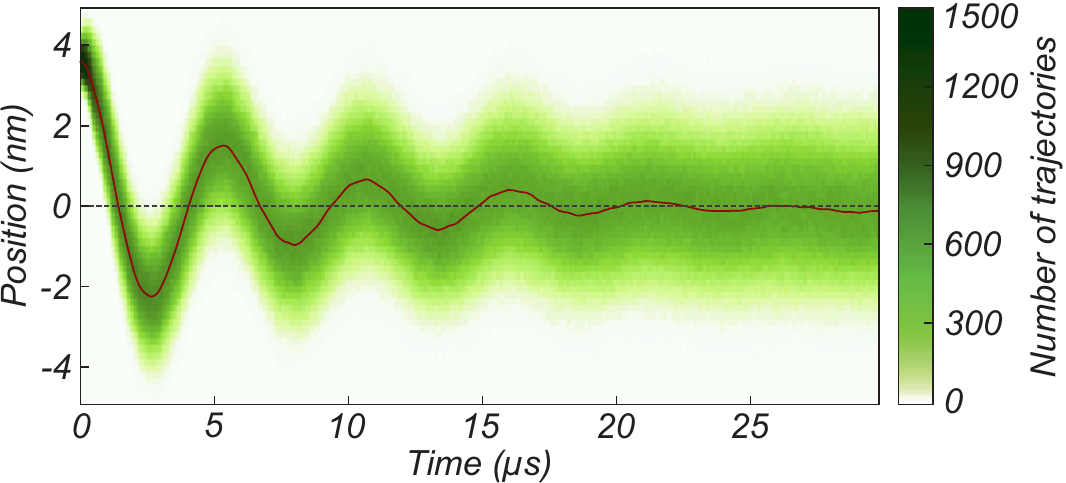}
\end{center}
\caption{Space-time correlations of the position of the nanowire extremity undergoing Brownian motion at $300\,\rm K$, measured using the QPD detectors (green trace). To compute this correlation a long lasting  signal (about $3\,\rm s$) from the QPD measurement apparatus is acquired (12-bits resolution) and calibrated in displacement.  1800 sections of $30\,\rm \mu s$ are singled out if they meet the two following criteria: start at a position around $3.8\,\rm nm$ from the equilibrium position and with a speed around $0\,\rm nm.s^{-1}$, with both selection windows set at the limit of resolution of the apparatus roughly given by the trace initial breadth. These criteria set a start point in phase space from where we observe its re-population under decoherence. The chosen criteria are therefore arbitrary and the figure serves only as an illustration for this measurement \cite{Li2010}. Note that an additional condition is added for the computation which is that no subsequent sections separated by less than $30\,\rm \mu s$ are retained in order to avoid parasitic correlations due to over-sampling. Eventually each of the chosen sections is digitized over a grid, and the sum of these grids give the correlation trace presented here in green, thus rebuilding in time domain part of the signal usually computed by a spectrum analyzer. The space average position (dark red solid line) is computed from the analysis of vertical slices of this trace.}
\label{fig:spacetime}
\end{figure}

The position of the nanowire  $\delta x (t)$ can be read out in real time  and recorded through the reflection or the transmission of the green laser light onto one of the two QPDs of the setup. It is then possible to compute the oscillator space-time correlations  $C_{\delta x}(\tau)=\langle \delta x(t)\delta x (t+\tau)\rangle$ as shown in Fig.  \ref{fig:spacetime}.\\
Since the readout laser is positioned on the NV location, the measured displacement $\delta x(t)$ also represents the transverse position of the NV defect $\xi(t)$.
Once again, we restrict our analysis to a mono-dimensional trajectory, perpendicular to the optical axis. The extrapolation to any other 3D spatial trajectory is straightforward once the PSFs $\Pi_i(\vv{r})$ are known.

Alternatively, the high frequency output of the QPD is sent to a Spectrum Analyzer (Agilent MXA) to compute the position spectrum. The final position spectrum showing several vibration eigenmodes of the nanowire in Fig. 1g was extracted from the Spectrum Analyzer raw measurement $S_{\rm V_{HF}}[\Omega]$ through:
\begin{equation}
S_{\rm x}[\Omega]=\Big({S}_{\rm V_{HF}}[\Omega]-{S}_{\rm dark_{HF}}[\Omega]\Big) \;\Bigg(\frac{G_{DC}[0]}{G_{HF}[\Omega]}\Bigg)^{2}\; \Bigg(\frac{\partial V_{DC}}{\partial x}\Bigg)^{-2}
\end{equation}
where $S_{\rm dark_{HF}}$ is the (dark) background noise of the QPD HF channel, $G_{DC}[\Omega]$ and $G_{HF}[\Omega]$ are the previously calibrated DC and HF frequency-dependent gains, and $\frac{\partial V_{DC}}{\partial x}$ is the DC-voltage-position conversion factor measured by making a DC image of the nanowire displaced with the piezoelectric stage. Typically, the laser shot noise emerges above the QPD dark noise level for optical powers larger than $\approx 100-200\,\rm\mu W.$

\begin{flushleft}
{\large \bf Electrostatic actuation}
\end{flushleft}

A conducting tungsten tip is piezo-positioned in the vicinity of the nanowire's extremity. A high voltage -- 50-fold amplification of an Agilent Arbitrary Waveform Generator typically $3\,\rm V$ signal -- is applied between the tip and the nanowire metallic holder. Other than this tip the whole setup is electrically connected to the holder and grounded.  The electrostatic force is the spatial gradient of the energy stored in the circuit formed by the nanowire and the electrostatic tip. For a tip positionned perpendicularly to the optical axis, the expression of the equivalent electrostatic force  applied on the apex of the nanowire can be written as:

%
\begin{equation}
{\bf F}({\bf r_{0}})= -\alpha V^{2}{\bf e}_{\rm x},
\end{equation}
where $V$ is the applied voltage difference and $\alpha$ contains all of the spatial and material dependence of the electrostatic description of the system.

According to classical beam theory, the static deflection of the nanowire is then in turn quadratic in voltage as shown on Fig. 2b:
\begin{equation}
|\delta{\bf x}|=\frac{\alpha V^{2}L^{3}}{3EI} = \kappa V^{2},
\end{equation}
where $L$ is the nanowire's length, $E$ its Young modulus, $I$ its moment of inertia, and $\kappa$ is a global electromechanical coefficient for the system.

The fit of the static deflection as a quadratic function of voltage shown in Fig. 2b gives an electromechanical coefficient of $\kappa_{\rm stat} = 8.53\cdot 10^{-11}\,\rm m.V^{-2}$.

The setup illustrated in Fig. 2a is used to simulate a Langevin force of arbitrary temperature with a white Gaussian voltage noise $\delta V$ of spectral density $\mathcal{S}_{V}[\Omega]$ and variance $\sigma_{V}$ added to a voltage offset $V_{0}$. This offset is chosen to be much greater than the force signal so that at first-order the dynamic force noise becomes linear in the noise voltage (at the cost of a high static force and thus a strong static deflection which is compensated by a displacement of the nanowire support): $\delta F\approx 2 \alpha V_0 \delta V$. The corresponding force spectral density becomes
\begin{equation}
\mathcal{S}_{\rm F}[\Omega]=4\alpha^{2}V_{0}^{2}\mathcal{S}_{V}[\Omega],
\end{equation}
and the effective temperature (the damping rate is unchanged):
\begin{equation}
T_{\rm eff}=T+\frac{2 \alpha^{2} V_{0}^{2}S_{V}}{M_{\rm eff} k_{B}\Gamma_{\rm m}}
\end{equation}

The signal generator bandwidth is chosen to fall short of the second mode frequency. Only the first eigenmode of the oscillator is sensible to this simulated high temperature. The flatness of the generated voltage signal was verified with a Spectrum Analyzer and its Gaussian character was also assessed through quadrature analysis for several frequency ranges.

The time-averaged spatial distribution of the NV center fluorescence  $\Phi(x_{0})$ is measured with an APD (Fig. 2a) under wide illumination field. When the nanowire is driven with an enhanced and spectrally white force, the r.m.s. position can be inferred from a Gaussian fit of the time-integrated fluorescence :
\begin{equation}
\Phi(x_{0}) = \Phi_{0}\int_{-\infty}^{+\infty} \Pi(x-x_0) P(x) {\rm d}x
=
\Phi_{0}\int_{-\infty}^{+\infty}  e^{-2 \frac{(x-x_{0})^{2}}{w_{0}^{2}}}\, \frac{e^{-\frac{x^{2}}{2\Delta x_{\rm th}^{2}}}}{\sqrt{2\pi \Delta x_{\rm th}^2}}{\rm d}x
=
  \frac{\Phi_{0}\,  e^{ - \frac{x_{0}^{2}}{w_0^2/4+ \Delta x_{\rm th}^{2}}}}{\sqrt{1+4\Delta x_{\rm th}^2/w_0^2}}
\end{equation}
where $w_{0}$ is the detection waist and $\Delta x_{th}$ the r.m.s. position.

\vspace{0.2cm}

The orange dots in Fig. 2f show the total fitted Gaussian width of the broadened fluorescence images. The red line shows the same width corrected for the detection waist to give $\Delta x_{\rm th}$. This is therefore a second method for calculating the electromechanical coefficient : $\kappa_{\rm dyn} = 6.70\cdot 10^{-11}\,\rm m.V^{-2}$ which  coincides reasonably well with the static characterisation. Yet another measurement of the r.m.s. position is given by the integration of the position  spectral density (green dots on Figure 2f), which is in agreement with the previous measurement as shown by the equal slopes between red and green dots at small effective temperatures on Fig. 2f. However in this measurement the position was measured from the QPD differential voltage  which has a limited linearity zone that the nanowire's motion exceeds when the temperature is increased, resulting in saturation effects.

\begin{flushleft}
{\large \bf Generic expressions of the initial autocorrelation strength}
\end{flushleft}
Using the spatial distribution of the Brownian motion
$$
P(x)=\frac{1}{\sqrt{2\pi \Delta x^2}} e^{-\frac{x^2}{2\Delta x^2}},
$$
and the displaced PSF,:
$$
\Pi_i(x)= e^{-2 (x-x_i)^2/w_0^2},
$$
we can estimate the initial bunching.
We have:
$$
G^{(2)}_{\rm osc}(\infty,x_1,x_2)\propto \int{dx P(x) \Pi_1(x)} \int{dx P(x) \Pi_2(x)}
$$
and
$$
G^{(2)}_{\rm osc}(0,x_1,x_2)\propto \int{dx P(x) \Pi_1(x) \Pi_2(x)}
$$
so the initial relative correlation strength gives, using $\tilde{\delta}_i\equiv x_i/w_0/\sqrt{2}$:
\begin{figure}[t]
\begin{center}
\includegraphics[width=0.9\linewidth]{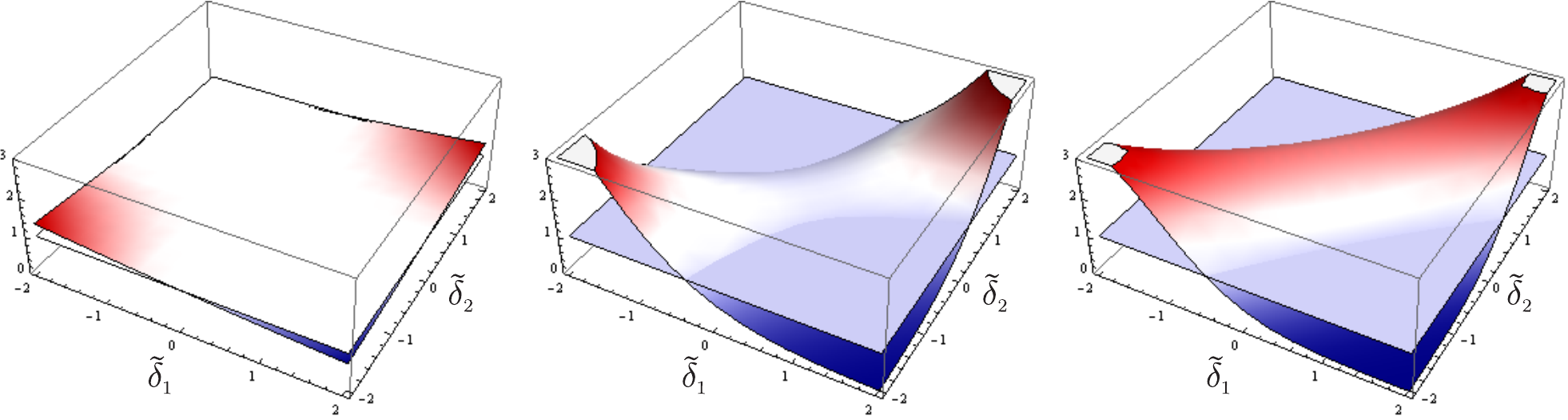}
\caption{Initial bunching/anti-bunching strength ${G^{(2)}_{\rm osc}(0,x_1,x_2)}/{G^{(2)}_{\rm osc}(\infty,x_1,x_2)}$  of motional origin obtained for varying start/stop detector positions $\tilde\delta_{1,2}=x_{1,2}/w_0/\sqrt{2}$ in the case of increasing oscillating amplitude $\theta_x=0.1, 0.5, 1.5$ (left to right).}
\label{FigS3}%
\end{center}
\end{figure}
\begin{equation}
\frac{G^{(2)}_{\rm osc}(0,x_1,x_2)}{G^{(2)}_{\rm osc}(\infty,x_1,x_2)}
=\frac{1+4\theta_x^2}{\sqrt{1+8\theta_x^2}}
\exp{\left(-\frac{8\theta_x^2}{(1+4\theta_x^2)(1+8\theta_x^2)}\left(2\theta_x^2(\tilde\delta_1-\tilde\delta_2)^2-\tilde\delta_1\tilde\delta_2\right)\right)}
\end{equation}
It is plotted in \ref{FigS3}, for varying vibration strength. In the centered case one obtains:
$$\frac{G^{(2)}_{\rm osc}(0,0,0)}{G^{(2)}_{\rm osc}(\infty,0,0)}
=\frac{1+4\theta_x^2}{\sqrt{1+8\theta_x^2}}>1
$$
which is  in good agreement with experimental results (see Fig.\,3d of the manuscript).

\vskip 2 cm

\begin{flushleft}
{\large \bf Simulations}
\end{flushleft}

As spin variables here are irrelevant, the NV center is modeled as a three-level system, keeping only one ground state $\ket{g}$ and one excited state $\ket{e}$ regardless of spin projection, and merging the whole $S=0$ levels system into one dark metastable state $\ket{m}$. The statistical population of these states will be denoted $\sigma_{g}$, $\sigma_{e}$ and $\sigma_{m}$ and normalized such that the sum of these three populations is always unity.
The fluorescence intensity is then proportional to the excited state population $\sigma_{e}$. Right after the emission of one photon, the state of the NV center is known to be $\ket{g}$. Thus, the intensity correlation can be computed from the evolution of $\sigma_{e}$ under the optical Bloch equations with the initial condition of  $\sigma_{g}=1, \sigma_{e,m}=0$ as $\sigma_{e}(t)$ then represents the probability of emitting a second photon. Any (slowly) time-dependent excitation intensity can be injected in the Bloch equations for this numerical resolution which allows us to build the intensity correlations for any type  of trajectory of the NV center in and out of the excitation volume. Additional spatial weighting is added to model the motion in and out of the start/stop collections volumes which may or may not coincide, depending on the type of measurement performed.\\

In adiabatic cases where the NV center's populations adapt almost immediately to the received optical power compared to the mechanical oscillation period, the resolution of the Bloch equations is not essential and the fluorescence, even modulated by the slow motion of the nanowire, can be computed either analytically for simple trajectories as sine-shaped ones or through statistical considerations for Brownian motion. However, in order to take into account the photodynamics of the NV center and investigate the non-adiabatic regimes using a simulation Runge-Kutta 4 algorithm to resolve optical Bloch equations.\\

Since the choice of trajectory is then completely arbitrary, this program can be fed with a simulated trajectory obtained with the convolution of a white noise are then used to build the full correlation.\\
Each of these simulated trajectory realizations is first given a weight depending on the presence or not of its starting point within a Gaussian window around the first detector's central position. This accounts for the start channel PSF. The subsequent probability of fluorescence through time is calculated by numerically solving the Bloch equation for 1 point out of 2 in the trajectory (as the Runge Kutta algorithm requires twice as many points as wanted in the output signal). Each of these fluorescence values is in turn pondered with a second weight to take into account the Gaussian window around the second detector's central position. The two detectors can therefore be moved separately.

\begin{flushleft}
{\large \bf Expansion of the photon autocorrelation}
\end{flushleft}

We show here that in case of broad illumination, the photon autocorrelation function can be expanded as:

\begin{equation}
\mathrm{g}^{(2)}\left(\tau,x_1,x_2\right)=\frac{\sigma_e(\tau)}{A_0}\sum_{j=0}^\infty{A_j \left(\frac{- C_\xi(\tau)}{w_0^2/2}\right)^j}
\label{eq.g2Aj}
\end{equation}
where  the $A_j$ coefficients are  delay independent and functions of the detector positions $x_i$ and oscillator spatial spreading $\theta_x$. One has:
$$
A_{2j}=\frac{1}{(2j)!} \sum_{n,m=0}^{\infty}\frac{(2n+2j)!(2m+2j)!}{(n+j)!(m+j)! }
\left(\frac{-2 x_1^2}{w_0^2}\right)^{n}
\left(\frac{-2 x_2^2}{w_0^2}\right)^{m}
\Lambda_{n,m}^{\rm even}
$$
and
$$
A_{2j+1}=\frac{-2x_1 x_2}{w_0^2 (2j+1)!} \sum_{n,m=0}^{\infty}\frac{(2n+2j+2)!(2m+2j+2)!}{(n+j+1)!(m+j+1)!}
\left(\frac{-2 x_1^2}{w_0^2}\right)^{n}
\left(\frac{-2 x_2^2}{w_0^2}\right)^{m}
\Lambda_{n,m}^{\rm odd}
$$
with

$$\Lambda_{n,m}^{\rm even}=\sum_{p,q=0}^{n,m}\frac{\left(\frac{\Delta x^2}{2 x_1^2}\right)^{p} \left(\frac{\Delta x^2}{2 x_2^2}\right)^{q} }{(2n-2p)! (2m-2q)! p! q!},$$

$$\Lambda_{n,m}^{\rm odd}=\sum_{p,q=0}^{n,m}
\frac{
\left(\frac{\Delta x^2}{2 x_1^2}\right)^{p} \left(\frac{\Delta x^2}{2 x_2^2}\right)^{q}
}{
(2n-2p+1)! (2m-2q+1)! p! q!}.$$

In case of broad spatial illumination, the measured fluorescence is modulated on short time scales by the photophysics of the NV center and on long time scales by the motion of the NV center in and out of the collection volumes.
Rewriting the expression introduced in the manuscript in case of broad illumination gives:
\begin{equation}
\mathrm{g}^{(2)}\left(\tau,x_{1},x_{2}\right)
=
\frac{
\sigma_e(\tau) \left\langle \Pi(\xi(t)-x_1) \Pi(\xi(t+\tau)-x_2)\right\rangle_T }
{\mathcal{N}}.
\end{equation}

We will now expand  in powers of $\xi(t)$ the two probabilities of detection centered respectively on $x_{1}$ and $x_{2}$, and in doing so we will only suppose that the PSFs  are expandable in power series. We will focus on the numerator $\mathrm{G}^{(2)}(\tau, x_1, x_2)$.
\begin{equation}
\mathrm{G}^{(2)}\left(\tau,x_1,x_2\right)=\sigma_e(\tau) \left\langle \sum_{n=0}^{\infty}\sum_{m=0}^{\infty}
\frac{d^{n}\Pi}{dx^{n}}\Big|_{x_1} \frac{d^{m}\Pi}{dx^{m}}\Big|_{x_2}
\frac{(\xi(t)-x_1)^{n}(\xi(t+\tau)-x_2)^{m}}{n!m!}\right\rangle_T
\end{equation}
\begin{equation}
\mathrm{G}^{(2)}\left(\tau,x_1,x_2\right)=\sigma_e(\tau) \sum_{n=0}^{\infty}\sum_{m=0}^{\infty} \frac{1}{n!m!} \frac{d^{n}\Pi}{dx^{n}}\Big|_{x_1} \frac{d^{m}\Pi}{dx^{m}}\Big|_{x_2} \sum_{p=0}^{n}\sum_{q=0}^{m} \begin{pmatrix}
n \\
p
\end{pmatrix}\begin{pmatrix}
m \\
q
\end{pmatrix}x_{1}^{n-p}x_{2}^{m-q}\left\langle\xi(t)^{p}\xi(t+\tau)^{q}\right\rangle
\label{eq.firstexpansion}
\end{equation}
where we have omitted the subscript $T$ in the time averaging. $\begin{pmatrix}
n \\
m
\end{pmatrix}$ denotes the $m$-combination from an ensemble of size $n$. Since $\xi(t)$ is a Gaussian distributed random variable, as visualized in Fig. 2 of the manuscript, a very useful property of Gaussian processes can be used to reduce $ \langle\xi(t)^{p}\xi(t+\tau)^{q}\rangle$ to a sum of small manageable terms.
This property holds as follows: for $n$ jointly Gaussian random variables $x_{\{i=0...n\}}$ \textit{whatever their auto- and cross-correlations}:
\begin{equation}
\begin{cases}
\langle x_{1}...x_{n}\rangle = \sum_{k\in \mathcal{C}_{\rm pairs}} \langle x_{k_{0}}x_{k_{1}}\rangle...\langle x_{k_{n/2-1}}x_{k_{n/2}}\rangle & \text{ if \textit{n} is even}\\
\langle x_{1}...x_{n}\rangle = 0 & \text{ if \textit{n} is odd}
\end{cases}
\end{equation}
where $\mathcal{C}_{\rm pairs}$ denotes the ensemble of pair configurations for the indexes $1...n$, where each index appears exactly one time and with no consideration of order.

\begin{figure}[ht!]
\begin{center}
\includegraphics[width=6cm]{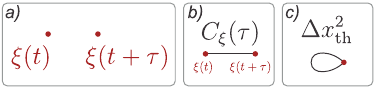}
\caption{Building blocks of the diagrammatic representation of terms. a) Dots are positions of the oscillator at different times. b) Lines that connect one dot to another represent autocorrelations. c) Loops connecting one dot are zero-delay autocorrelation.}
\label{fig:DiagExpl}
\end{center}
\end{figure}

This property therefore holds for our random variables $\{\xi(t),...,\xi(t),\xi(t+\tau),...,\xi(t+\tau)\}$ and enables the development of $\langle\xi^{p}(t)\xi^{q}(t+\tau)\rangle$ into terms of the form: $C_{\xi}^{i}(\tau) \Delta x_{\rm th}^{2j}$. In order to amount the contribution of each term to the total correlation, a diagrammatic picture of the problem was built upon the principles illustrated in \ref{fig:DiagExpl}. In this picture, a vertex represents the position at a time $t$, $\xi(t)$, and a line joining two vertices represents a correlation between these two positions, that is, autocorrelations of the oscillator position. A line can also join one vertex to itself making a loop which thus represents the zero-delay autocorrelation.

\begin{figure}[ht!]
\begin{center}
\includegraphics[width=12cm]{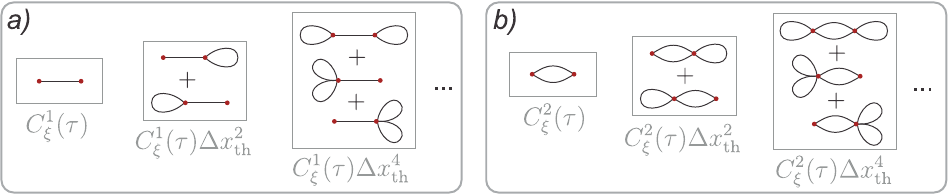}
\caption{Example of terms reprsented with the diagrammatic picture explained above. a) Non exhaustive list of diagrams contributing to the first power of correlation in the total expansion. b) Non exhaustive list of diagrams contributing to the second power of correlation in the total expansion. Each of these diagram is weighed by a factor calculating in the following.}
\label{fig:DiagExam}
\end{center}
\end{figure}

Some examples of this representation are given in \ref{fig:DiagExam} for the first two orders in correlation of the expansion.

Counting all the possible ways to obtain a term where the correlation appears at a fixed power $j$ : $C_{\xi}^{j}(\tau)\Delta x_{\rm th}^{(p+q)-2j}$ from the term $\langle\xi^{n}(t)\xi^{m}(t+\tau)\rangle$ with no consideration of order, we obtain that the possibilities are numbered as explained in \ref{fig:DiagDegen}:
\begin{figure}[ht!]
\begin{center}
\includegraphics[width=17cm]{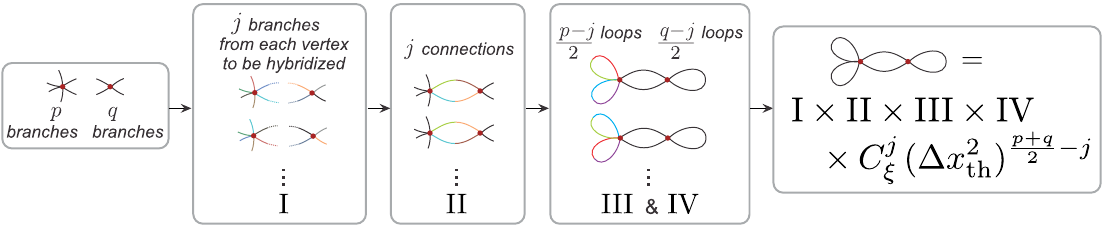}
\caption{Diagrammatic explanation for the calculation of the degeneracy factor of each diagram representing a term $C_{\xi}^{j}(\tau)\Delta x_{\rm th}^{(p+q)-2j}$ (complementary to equation \ref{eq:degen}). Starting from the knowledge that each vertex must be connected $p$ ($q$) times and that $j$ lines must connect them, there is a number I of possibilities to select $j$ branches from each vertex to build interconnections. The lines are colored to identify them when it is relevant. After this selection of $j$ branches per vertex, there is still a number II matching possibilities. Then the remaining branches are connected to other branches from the same vertex, which makes up another degeneracy factor III for one vertex and IV for the other.}
\label{fig:DiagDegen}
\end{center}
\end{figure}
\begin{equation}
\underbrace{\begin{pmatrix}
p \\
j
\end{pmatrix}\begin{pmatrix}
q \\
j
\end{pmatrix}}_{\displaystyle \rm I}
\underbrace{\begin{matrix}
\,^{\,} \\
\,_{\,}
\end{matrix}j!}_{\displaystyle \rm II}
\underbrace{\frac{(p-j)!}{(\frac{p-j}{2})!2^{\frac{p-j}{2}}}}_{\displaystyle \rm III} \underbrace{\frac{(q-j)!}{(\frac{q-j}{2})!2^{\frac{q-j}{2}}}}_{\displaystyle \rm IV}
\label{eq:degen}
\end{equation}

I is the number of ways to select $j$ elements from each ensemble of size $p$ and $q$ of variables $\xi(t)$ and $\xi(t+\tau)$ to produce a hybrid pair $C_{\xi}^{j}(\tau)$. Once $j$ elements from each group are selected, there are still $j!$ ways to combine them to make $C_{\xi}^{j}(\tau)$ which is represented by the factor II. Identically, the remaining $p-j$ and $q-j$ elements in each group can be associated two by two in different ways which results in degeneracy factor III and IV. Note that $j$, $p$ and $q$ have the same parity: the terms that would not respect this condition do not exist in the development. After simplifications one finally obtains for $p$ and $q$ of the same parity:
\begin{equation}
\left\langle\xi(t)^{p}\xi(t+\tau)^{q}\right\rangle= \underbrace{\sum_{j=0}^{{\rm min}(p,q)}}_{\text{j of p,q parity}} C_{\xi}^{j}(\tau) (\Delta x_{th}^{2})^{\frac{p+q}{2}-j} \frac{p!q!}{2^{\frac{p+q}{2}-j}j!(\frac{p-j}{2})!(\frac{q-j}{2})!}
\end{equation}
and $\left\langle\xi(t)^{p}\xi(t+\tau)^{q}\right\rangle=0$ if $p$ and $q$ have different parities.

\vspace{0.5cm}

Now assuming the collection volumes $\Pi_{1}$ and $\Pi_{2}$ to be Gaussian  (more precisely, TEM00 profiles at waist), centered in $x_{1}$ and $x_{2}$, and normalized so that the maximum detection is unity,
\begin{equation}
\Pi_{1}(x)=e^{-\frac{2(x-x_{1})^{2}}{w_{0}^{2}}},\quad\Pi_{2}(x)=e^{-\frac{2(x-x_{2})^{2}}{w_{0}^{2}}},
\end{equation}
one can finally compute from equation (\ref{eq.firstexpansion}), using a serial expansion of $e^{-x^2}$, the total development of the numerator of the autocorrelation function. Finally, it is of special interest to rewrite this development as a sum over the powers of $C_{\xi}^{j}$, which results after rearranging the terms in the numerator of equation \ref{eq.g2Aj}.\\
Once the numerator is expanded, we can point out that at infinite delay $\tau$, the mechanical autoccorelation function converges towards 0, so that we are only left with the $A_0$ term of the series expansion. As a consequence, the denominator equals $A_0$.

Let us mention here that very efficient algorithms based on cumulants can be exploited to directly compute the overall oscillatory autocorrelation function $\mathrm{G}^{(2)}_{\rm osc}\left(\tau,x_{1},x_{2}\right)$. \cite{Pistolesi2014}.

\begin{figure}[h!]
\begin{center}
\includegraphics[width=0.9\linewidth]{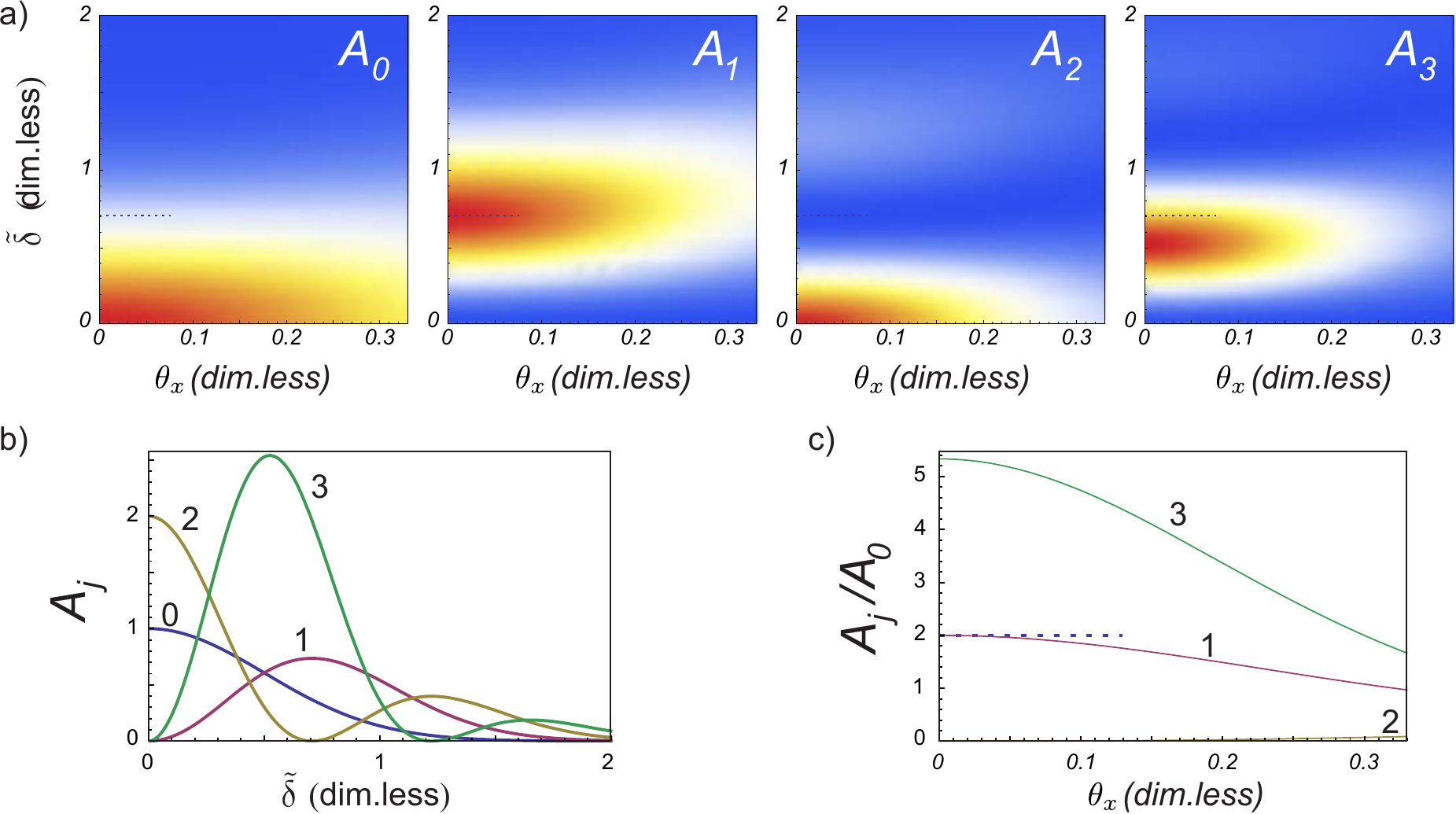}
\caption{$A_j$ expansion in the symmetric case $g^{(2)}\left(\tau,\delta,-\delta\right)$  a): Strength of the $A_j$ terms calculated from the series expansion ((\ref{eq.A2j}), (\ref{eq.A2j+1})), as a function of  the normalized temperature $\theta_x$ and detector position $\tilde \delta$ showing the  cancelation of $A_2$ in $ \tilde\delta=1/\sqrt{2}$ at small excitation.
b) cross section for $\theta_x=0.001$ c) evolution of $A_j/A_0$ for $\tilde\delta=1/\sqrt{2}$ as a function of $\theta_x$. This validates the approximated expansion and at low drive excitation.}
\label{FigS1}%
\end{center}
\end{figure}

\begin{flushleft}
{\large \bf Symmetric case $\mathrm{g}^{(2)}\left(\tau,\delta,-\delta\right)$}
\end{flushleft}
We study here the case where both detectors are symmetrically positioned, monitoring each side of the trajectory: $x_1=-x_2= \delta$. We introduce the normalized detector position and oscillator temperature:
$$ \tilde \delta=\frac{\delta}{w_0/\sqrt{2}}{\rm\,\,\ and \,\ } \theta_x=\frac{\Delta x}{w_0}.$$
One then obtains:

\begin{equation}
A_{2j}(\delta,-\delta)=
\frac{1}{(2j)!} \left(\sum_{n=0}^{\infty}\frac{(2n+2j)!}{(n+j)! }
\left(-\tilde \delta^2\right)^{n}
\sum_{p=0}^{n}\frac{
\left({\theta_x^2}/{\tilde \delta^2}\right)^{p}
}{(2n-2p)!  p! }\right)^2
\label{eq.A2j}
\end{equation}
and
\begin{equation}
A_{2j+1}(\delta,-\delta)=
\frac{\tilde \delta^2}{ (2j+1)!} \left(\sum_{n=0}^{\infty}\frac{(2n+2j+2)!}{(n+j+1)!}
\left(-\tilde \delta^2\right)^{n}
\sum_{p=0}^{n}
\frac{
\left({\theta_x^2}/{\tilde \delta^2}\right)^{p}
}{
(2n-2p+1)! p! }\right)^2
\label{eq.A2j+1}
\end{equation}
For small  oscillation amplitudes, $\theta\ll 1$, one can restrict the second finite sum to the power $p=0$. The $A_j$ coefficients are then temperature independent and one has:

$$A_{0}(\delta,-\delta)\approx
 \left(\sum_{n=0}^{\infty}\frac{\left(-\tilde \delta^2\right)^{n}}{n! }
\right)^2 = e^{-2\tilde \delta^2}=e^{- \frac{4\delta^2}{w_0^2}},
$$

$$A_{1}(\delta,-\delta)\approx
\tilde \delta^2 \left(\sum_{n=0}^{\infty} 2 \frac{\left(-\tilde\delta^2\right)^{n}}{n! }
\right)^2 =4\tilde\delta^2 e^{-2\tilde\delta^2},
$$

$$A_{2}(\delta,-\delta)\approx
\frac{1}{2} \left(\sum_{n=0}^{\infty} 2 \frac{(2n+2 -1)\left(-\tilde\delta^2\right)^{n}}{n! }
\right)^2 =\frac{1}{2} \left( 4 \frac{d}{d(-\tilde\delta^2)} {(-\tilde\delta^2 e^{-\tilde\delta^2})} -2 {e^{-\tilde\delta^2}} \right)^2 =2(1-2\tilde\delta^2)^2 e^{-2\tilde\delta^2}.
$$
From that it is clear that $A_2$  is suppressed for $\tilde\delta=\tilde\delta^\star\equiv 1/\sqrt{2}$, i.e. $\delta=w_0/2$. One then obtains:
 $$A_{0}(\delta^\star,-\delta^\star)\approx1/e\,\, , \,  A_{1}(\delta^\star,-\delta^\star)\approx2/e$$
For higher spatial spreading $\theta_x$, the previous simplification is not possible.
The dependency of the $A_j$ coefficient in $\delta$ and $\theta_x$ is shown in \ref{FigS1}. It is still possible to cancel the $A_2$ term by choosing the proper detector displacement. The numerical quantities have been calculated up to $\theta_x=0.35$, beyond which numerical convergence is not achieved.\\

At low excitation, one then obtains:
$$ g^{(2)}(\tau,\delta^\star,-\delta^\star)\approx \sigma_e(\tau)\left(1-4 \frac{C_\xi(\tau)}{w_0^2}+...\right).$$
The mechanical spectrum is then convoluted with the response function corresponding to the Fourier transform of $\sigma_e(\tau)$, which behaves as a low pass filter of cutoff frequency ($1/\tau_c$). The number of detected photons in each time bin after an integration time $T$ is $\Phi_1\Phi_2 \tau_{\rm bin} T$, where $\tau_{\rm bin}$ is the bin time. For a Poissonian distribution, assuming identical fluxes on each APD, the equivalent sensitivity of the apparatus is then given by
$$C_\xi^{\rm min}=\frac{w_0^2}{4 \Phi \sqrt{\tau_{\rm bin}}}\approx (6\,\rm nm)^2/\sqrt{Hz}$$
for a photon flux of $10^6\,\rm cps$ (300\,fW at 700\,nm) and a time bin of $\tau_{\rm bin}=1\,\mu\rm s$. This value can be compared  favourably to the thermal noise of a 1\,MHz ($M_{\rm eff}=10^{-20}\,\rm kg$) suspended carbon nanotube: $\Delta x_{\rm th}=\sqrt{ k_b T/M\Omega_{\rm m}^2}\approx 102\,\rm nm$ at 300\,K and 12\,nm at 4\,K. In the article,  all the measurements have been carried out at a 1.6\,ns bin time to resolve the anti-bunching signatures - at a price of a longer acquisition time (1 hour for Fig. 4d) - which is not necessary to resolve the oscillating bunching pattern of motional origin.

\begin{figure}[t!]
\begin{center}
\includegraphics[width=0.9\linewidth]{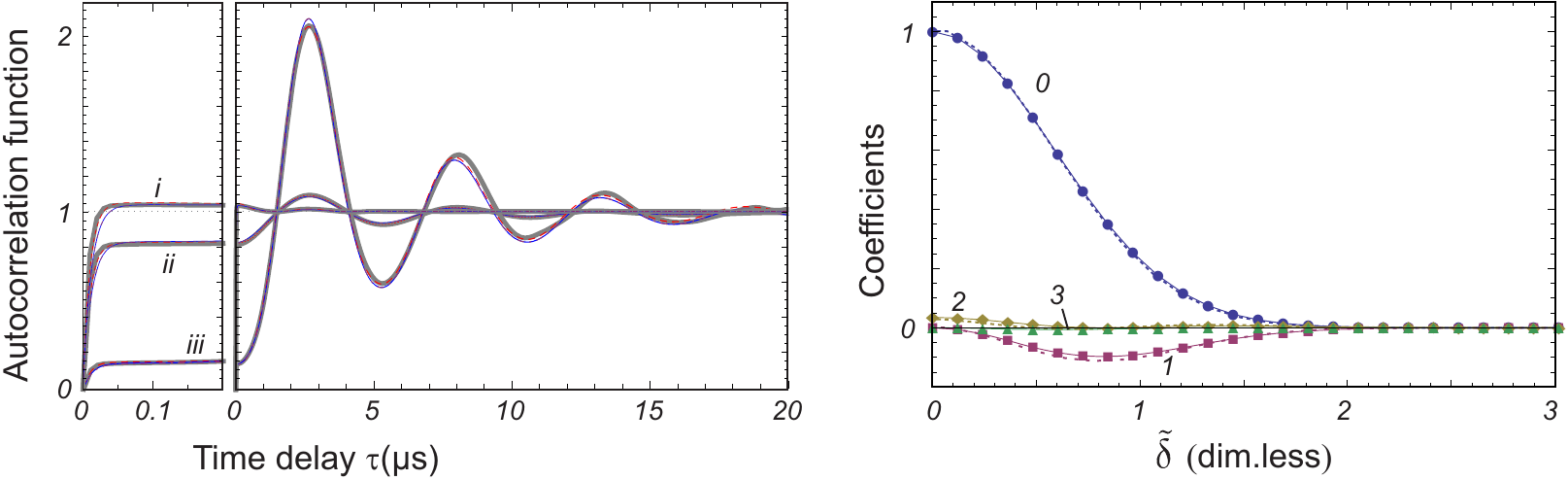}
\caption{Comparison of the numerical simulations to the $A_j$ expansion. Left: numerical simulations  of $g^{(2)}\left(\tau,\delta,-\delta\right)$ for $\theta_x=\Delta x_{\rm th}/w_0=0.3$ (gray) for varying detector positions $\tilde\delta=0, 0.64, 1.95$ from {\it i} to {\it iii}. The data are fitted with an expansion up to the $4^{th}$ order in $c(\tau)$ (equation (\ref{eq.alphaj})) (red dashed traces). Also shown in blue is expression (\ref{eq.g2Aj}) where the $A_j$ coefficients were computed using expressions (\ref{eq.A2j},\ref{eq.A2j+1}). Right:  comparison of the fitted $\alpha_j (\delta,-\delta)/ \alpha_0(0,0)$ coefficients (connected symbols) and the respective coefficients $A_j(\delta,-\delta) (2\theta_x)^{-j}$ (dashed lines), which have been numerically evaluated using the 500 first terms of the series. The good agreement achieved validates the expansion (\ref{eq.g2Aj}), even for strong driving.}
\label{FigS2}%
\end{center}
\end{figure}

\begin{flushleft}
{\large \bf Comparison between simulations and $A_j$ expansion}\\
\end{flushleft}
Here we verify that the expansion in powers of the oscillator autocorrelation function $C_\xi(\tau)$ gives results in agreement with the direct numerical simulation of the oscillating NV fluorescence properties, introduced above. The comparison is presented here for a given excitation  $\Delta x_{\rm th}/w_0=0.3$ and for increasing distance between the detection areas, in the symmetric case: $x_1=-x_2=\delta$.
The results of numerical simulation are shown in \ref{FigS2} where they are also compared to the expansion up to the $4^{th}$ order in $C_\xi(\tau)$, using expressions (\ref{eq.A2j}), (\ref{eq.A2j+1}) and similar photophysical and mechanical parameters as the one employed in the numerical simulation.
The numerical simulations are also fitted with:
\begin{equation}
g^{(2)}\left(\tau,\delta,-\delta\right)=\frac{\sigma_e(\tau)}{\alpha_0}\sum_{j=0}
^4{\alpha_j \left(-\frac{C_\xi(\tau)}{\Delta x_{\rm th}^2}\right)^j},
\label{eq.alphaj}
\end{equation}
the only fitting parameters being the $\alpha_j$.  They are compared to the coefficients $A_j \left(\frac{w_0^2/2}{\Delta x_{\rm th}^2}\right)^j$ in \ref{FigS2} were a very good agreement is found. The slight deviation visible in the temporal traces obtained at large detector separation $\delta$ might be due to the rarefaction of the detection events which requires larger computational strength, and to the fact that the expansion to the 4th order may become limited for perfectly describing this situation.

\begin{flushleft}
{\large \bf Centered case: $g^{(2)}(\tau,0,0)$}
\end{flushleft}

Notice that $A_{2j+1}\rightarrow 0$ when $x_i\rightarrow 0$. For  even coefficients, since  $\delta$  appears at the power $(\delta^2)^{n-p}$, the only remaining coefficient in the second sum is $p=n$ which permits to express:

 $$A_{2j}(0,0)=
\frac{1}{(2j)!} \left(\sum_{n=0}^{\infty}\frac{(2n+2j)!}{(n+j)! n!}
\left(-\theta_x^2\right)^{n}
\right)^2.
$$
For simplicity, data were fitted with a normalized autocorrelation function:  $c(\tau)\equiv C_\xi(\tau)/\Delta x^2$:
$$\mathrm{g}^{(2)}_{\rm exp}\left(\tau,\tilde\delta,-\tilde\delta\right )=\sigma_e(\tau)\frac{1}{\alpha_0}\sum_{j=0}^\infty{\alpha_j\, c(\tau)^j}.
$$
The connection between both coefficients is then:
$$
\alpha_j=A_j\left( \frac{\Delta x^2}{w_0^2/2}\right)^j= A_j\left(2\theta_x^2 \right)^j
$$

\begin{flushleft}
{\large \bf Space-Time diffusion case: $g^{(2)}\left(\tau,0,\delta\right)$}
\end{flushleft}
In that case, using the same normalized quantities and following the same reasoning gives:
 $$A_{2j}(0,\delta)=
\frac{1}{(2j)!} \left(\sum_{n=0}^{\infty}\frac{(2n+2j)!}{(n+j)! n!}
\left(-\theta_x^2\right)^{n}
\right)
\left(\sum_{n=0}^{\infty}\frac{(2n+2j)!}{(n+j)! }
\left(-\tilde\delta^2\right)^{n}
\sum_{p=0}^{n}\frac{
\left({\theta_x^2}/{\tilde\delta^2}\right)^{p}
}{(2n-2p)!  p! }\right)
$$
and $A_{2j+1}(0,\delta)=0$

In both situations ($g^{(2)}\left(\tau,0,0\right)$ and $g^{(2)}\left(\tau,0,\delta\right)$) the nullity of odd coefficients $A_{2j+1}$ renders the motion dependent part of the photon autocorrelation functions $g^{(2)}_{\rm osc}(\tau,0,0)$ an even function of $C_\xi(\tau)$. Thus it does not oscillate above and below its limit at infinite delay.  This is particularly visible in Fig.\,3 and 4a and also gives rise to a contrast inversion which is visible in Fig. 4a, 4c.


\end{document}